\documentclass[final,tnotealph]{aipproc}
\layoutstyle{8x11single}
\usepackage{amsmath,amssymb}
\def\apj{Astrophys.\ J.\ }
\def\apjl{Astrophys.\ J.\ Lett.\ }
\def\apjs{Astrophys.\ J.\ Suppl.\ }
\def\mnras{Mon.\ Not.\ R.\ Astron.\ Soc.\ }
\begin{document}
%
\title{
Low $T/|W|$ dynamical instabilities in differentially rotating stars:
Diagnosis with canonical angular momentum}

%
\author{Motoyuki Saijo}{
  address={School of Mathematics, University of Southampton,
Southampton SO17 1BJ, United Kingdom}
}

\author{Shin'ichirou Yoshida}{
  address={Department of Physics, Florida Atlantic University, 
Boca Raton, FL 33431, USA}
}

\begin{abstract}
We study the nature of non-axisymmetric dynamical instabilities in  
differentially rotating stars with both linear eigenmode analysis and
hydrodynamic simulations in Newtonian gravity.  We especially
investigate the following three types of instability; the one-armed
spiral instability, the low $T/|W|$ bar instability, and the high
$T/|W|$ bar instability, where $T$ is the rotational kinetic energy
and $W$ is the gravitational potential energy.  The nature of the
dynamical instabilities is clarified by using a canonical angular
momentum as a diagnostic.  We find that the one-armed spiral and the
low $T/|W|$ bar instabilities occur around the corotation radius, and
they grow through the inflow of canonical angular momentum around the
corotation radius.  The result is a clear contrast to that of a
classical dynamical bar instability in high $T/|W|$.  We also discuss
the feature of gravitational waves generated from these three types of
instability. 
\end{abstract}
\maketitle

\section{Introduction}
Stars in nature are usually rotating and may be subject to
non-axisymmetric rotational instabilities.  An analytically exact
treatment of these instabilities in linearized theory exists only for
incompressible equilibrium fluids in Newtonian gravity
\citep[e.g.,][]{Chandra69}.  For these configurations, global
rotational instabilities may arise from non-radial toroidal modes
$e^{im\varphi}$ (where $m=\pm 1,\pm 2, \dots$ and $\varphi$ is the
azimuthal angle).

For sufficiently rapid rotation, the $m=2$ bar mode becomes either
{\em secularly} or {\em dynamically} unstable.  The onset of
instability can typically be marked by a critical value of the
dimensionless parameter $\beta \equiv T/|W|$, where $T$ is the
rotational kinetic energy and $W$ the gravitational potential
energy.  Uniformly rotating, incompressible stars in Newtonian theory
are secularly unstable to bar-mode formation when $\beta \gtrsim
\beta_{\rm sec} \simeq 0.14$.  This instability can grow only in the
presence of some dissipative mechanism, like viscosity or
gravitational radiation, and the associated growth time-scale is the
dissipative time-scale, which is usually much longer than the
dynamical time-scale of the system.  By contrast, a dynamical
instability to bar formation sets in when $\beta \gtrsim \beta_{\rm
  dyn} \simeq 0.27$.  This instability is present independent of any
dissipative mechanism, and the growth time is the hydrodynamic
time-scale. 

In addition to the classical dynamical instability mentioned above,
there have been several studies indicating that a dynamical
instability of the rotating stars occurs at low $T/|W|$, which is far
below the classical criterion of dynamical instability in Newtonian
gravity.  \citet{TH90} find in the self-gravitating ring and tori that
an $m=2$ dynamical instability occurs around $T/|W| \sim 0.16$ in the
lowest case in the centrally condensed configurations.  For rotating
stellar models, \citet*{SKE} find that $m=2$ dynamical instability
occurs around $T/|W| \sim O(10^{-2})$ for a high degree ($\Omega_{\rm
  c} / \Omega_{\rm s} \approx 10$) of differential rotation.  Note
that $\Omega_{\rm c}$ and $\Omega_{\rm s}$ are the angular velocity at
the centre and at the equatorial surface, respectively.
\citet{CNLB01} found dynamical $m=1$ instability around $T/|W| \sim
0.09$ in the $N = 3.33$ polytropic ``toroidal'' star with a high degree
($\Omega_{\rm c} / \Omega_{\rm s} = 26$) of differential rotation, and
\citet*{SBS03} extended the results of dynamical $m=1$ instability to
the cases with polytropic index $N \gtrsim 2.5$ and $\Omega_{\rm c} /
\Omega_{\rm s} \gtrsim 10$.

Computation of the onset of the dynamical instability, as well as the  
subsequent evolution of an unstable star, performed in a fully
nonlinear hydrodynamic simulation in Newtonian gravity have shown
that $\beta_{\rm dyn}$ depends only very weakly on the stiffness of
the equation of state.  Once a bar has developed, the formation of a
two-arm spiral plays an important role in redistributing the angular
momentum and forming a core-halo structure.  $\beta_{\rm dyn}$ are
smaller for stars with high degree of differential rotation
\citep{TH90, SKE, PDD}.  Simulations in relativistic gravitation
\citep{SBS00,SSBS} have shown that $\beta_{\rm dyn}$ decreases with
the compaction of the star, indicating that relativistic gravitation
enhances the bar mode instability. 

One of the remarkable features of these low $T/|W|$ instabilities is
an appearance of the corotation modes.  As it is pointed out by
\citet*{WAJ05} the low $T/|W|$ unstable oscillation of bar-typed one
found by \citet{SKE} has a corotation point.  Here corotation
means the pattern speed of oscillation in the azimuthal direction
coincides with a local rotational angular velocity of the star.  It is
well-known in the context of stellar or gaseous disk system that the
corotation of oscillation may lead to instabilities.  For instance,
there have been several density wave models proposed to explain spiral
pattern in galaxies, in which wave amplification at the corotation
radius of spiral pattern is a key issue.  Another example of
importance of corotation is found in the theory of thick disk (torus)
around black holes.  Initiated by a discovery of a dynamical
instability of geometrically thick disk by \citet{PP84}.
Instabilities of these systems are thought not to be unique in their
origin and in their characteristics.  Some seem to be related to local
shear of flow and to share a nature with Kelvin-Helmholtz instability.
Others may be related to corotation of oscillation modes with averaged
flow on which the oscillation is present.  The mechanisms of
instabilities by corotation, however, seem not unique.  As is
reminiscent to ``Landau amplification'' of plasma wave, a resonant
interaction of corotating wave with the background flow (in the case
of Landau amplification, background flow is that of charged particles)
may amplify the wave, by direct pumping of energy from background flow
to the oscillation.  The other may be an overreflection of waves at
the corotation which may be seen in waves propagating towards shear
layer. 

The main purpose of this paper, in contrast to the preceding studies
of this issue, is to investigate the nature of low $T/|W|$ dynamical
instabilities, especially to study the qualitative difference of them
from the classical bar instability.  As is mentioned above, recent
studies have shown that dynamical instabilities are possible for
different region of the parameter space of rotating stars.  Observing
the existence of dynamical instabilities whose critical $T/|W|$ value
are well below the classical criterion of bar instability, it is
natural to raise a question on whether these two types, ``high $T/|W|$''
and ``low $T/|W|$'', of dynamical instability are categorized in the
same type of dynamical instability or not.

Our study is done with both eigenmode analysis and hydrodynamical
analysis.  A non-linear hydrodynamical simulation is indispensable for
investigation of evolutionary process and final outcome of
instability, such as bar formation and spiral structure formation.
The nature of instability as a source of gravitational wave, which
interests us most, is only accessible through non-linear
hydrodynamical computations.  On the other hand, a linear eigenmode
analysis is in general easier to approach the dynamical instability of
a given equilibrium and it may be helpful to have physical insight on
the mechanism and the origin of the instability.  Therefore, a linear
eigenmode analysis and a non-linear simulation are complementary to
each other and they both help us to understand the nature of dynamical
instability.

As a simplified system mimicking the physical nature of the
differentially rotating fluid, we choose to study self-gravitating
cylinder models.  They have been used to study general dynamical
nature of such gaseous masses as stars, accretion disks and of stellar
system as galaxies.  Although there is no infinite-length cylinder in
the real world, it is expected to share some qualitative similarities
with realistic astrophysical objects \citep[e.g.][]{Ostriker65}.  
Especially it has served as a useful model to investigate secular and 
dynamical instabilities of rotating masses.  These works took advantage 
of a simple configuration of a cylinder compared with a spheroid.

This paper is organized as follows.  In the section: Dynamical
instabilities in differentially rotating stars, we present our 
hydrodynamical results of dynamical one-armed spiral and dynamical
bar instabilities.  We present our diagnosis of dynamical $m=1$ and
$m=2$ instabilities by using a canonical angular momentum in the
section: Canonical angular momentum to diagnose dynamical instability, 
and summarize our findings in the section: Summary and discussion.
Throughout this paper we use gravitational units with $G = 1$.  Latin
indices run over spatial coordinates.  A more detailed discussion is
presented in \citet{SY06}.

\section{Dynamical Instabilities in Differentially Rotating Stars}
\label{sec:Nhydro}

We explain three types of dynamical instabilities in differentially 
rotating stars based on non-linear hydrodynamical computations.  We
assume a polytropic equation of state only to construct an equilibrium
star as 
\begin{equation}
P = \kappa \rho^{1+1/N},
\end{equation}
where $P$ is a pressure, $\rho$ a rest-mass density, $\kappa$ a
constant, $N$ the polytropic index.  We also assume the
``$j$-constant'' rotation law, which has an exact meaning in the limit
of $d \rightarrow 0$, of the rotating stars 
\begin{equation}
\Omega = \frac{j_{0}}{d^{2} + \varpi^{2}},
\label{eqn:omega}
\end{equation}
where $\Omega$ is the angular velocity, $j_{0}$ a constant parameter
with units of specific angular momentum, and $\varpi$ the cylindrical
radius.  The parameter $d$ determines the length scale over which
$\Omega$ changes; uniform rotation is achieved in the limit $d
\rightarrow \infty$, with keeping the ratio $j_0/d^2$ finite.  We
choose the same data sets as \citet{SBS03} for investigating low
$T/|W|$ dynamical instabilities in differentially rotating stars
(models I and II in Table~\ref{tab:initial} corresponds to Tables II
and I of \citet{SBS03}, respectively).  We also construct an
equilibrium star with weak differential rotation in high $T/|W|$,
which excites the standard dynamical bar instability,
\citep[e.g.,][]{Chandra69}.  The characteristic parameters are
summarized in Table~\ref{tab:initial}.

\begin{table}
\begin{tabular}{ccccccccc}
\hline
Model &
$N$ \footnote[1]{$N$: Polytropic index} &
$d / R_{\rm eq}$ \footnote[2]{$R_{\rm eq}$: Equatorial radius} &
$R_{\rm pl} / R_{\rm eq}$ \footnote[3]{$R_{\rm pl}$: Polar radius} &
$\Omega_{\rm c} / \Omega_{\rm s}$ \footnote[4]
{$\Omega_{\rm c}$: Central angular velocity;
$\Omega_{\rm s}$: Equatorial surface angular velocity} &
$\rho_{\rm c} / \rho_{\rm max}$ \footnote[5]
{$\rho_{\rm c}$: Central mass density; 
$\rho_{\rm max}$: Maximum mass density} &
$R_{\rm maxd}/R_{\rm eq}$ \footnote[6]
{$R_{\rm maxd}$: Radius of maximum density} &
$T/|W|$ \footnote[7]{$T$: Rotational kinetic energy;
$W$: Gravitational binding energy} &
Dominant unstable mode
\\
\hline
I & $3.33$ &
$0.20$ & $0.413$ & $26.0$ & $0.491$ & $0.198$ &
$0.146$ & $m=1$
\\
II & $1.00$ &
$0.20$ & $0.250$ & $26.0$ & $0.160$ & $0.383$ &
$0.119$ & $m=2$
\\
III & $1.00$ &
$1.00$ & $0.250$ & $2.0$ & $0.837$ & $0.388$ &
$0.277$ & $m=2$
\\
\hline
\end{tabular}
\caption
{Three differentially rotating equilibrium stars that trigger
  dynamical instability}
\label{tab:initial}
\end{table}

To enhance any $m=1$ or $m=2$ instability, we disturb the initial
equilibrium mass density $\rho_{\rm eq}$ by a non-axisymmetric
perturbation according to 
\begin{equation}
\rho = \rho_{\rm eq}
\left( 1 +
  \delta^{(1)} \frac{x+y}{R_{\rm eq}} +
  \delta^{(2)} \frac{x^{2}-y^{2}}{R_{\rm eq}^{2}}
\right),
\label{eqn:DPerturb}
\end{equation}
where $R_{\rm eq}$ is the equatorial radius, with $\delta^{(1)} =
\delta^{(2)} \approx 1.7 - 2.8 \times 10^{-3}$ in all our simulations.
We also introduce ``dipole'' $D$ and ``quadrupole'' $Q$ diagnostics to
monitor the development of $m=1$ and $m=2$ modes as 
\begin{eqnarray}
D &=& \left< e^{i m \varphi} \right>_{m=1} =
\frac{1}{M} \int \rho \frac{x + i y}{\varpi} dV,
\\
Q &=& \left< e^{i m \varphi} \right>_{m=2} =
\frac{1}{M} \int \rho \frac{(x^{2}-y^{2}) + i (2 x y)}{\varpi^{2}} dV,
\end{eqnarray}
respectively.

We compute approximate gravitational waveforms by using the quadrupole  
formula.  In the radiation zone, gravitational waves can be described
by a transverse-traceless, perturbed metric $h_{ij}^{TT}$ with respect
to flat spacetime.  In the quadrupole formula, $h_{ij}^{TT}$ is found
from 
\begin{equation}
h_{ij}^{TT}= \frac{2}{r} \frac{d^{2}}{d t^{2}} I_{ij}^{TT},
\label{eqn:wave1}
\end{equation}
where $r$ is the distance to the source, $I_{ij}$ the quadrupole
moment of the mass distribution, and where $TT$ denotes the
transverse-traceless projection.  Choosing the direction of the wave
propagation to be along the rotational axis ($z$-axis), the two
polarization modes of gravitational waves can be determined from
\begin{equation}
h_{+} \equiv \frac{1}{2} (h_{xx}^{TT} - h_{yy}^{TT}),
\quad 
h_{\times} \equiv h_{xy}^{TT}.
\end{equation}
For observers along the rotation axis, we thus have
\begin{eqnarray}
\frac{r h_{+}}{M} &=&
\frac{1}{2 M} \frac{d^{2}}{d t^{2}} ( I_{xx}^{TT} - I_{yy}^{TT}),
\label{h+} 
\\
\frac{r h_{\times}}{M} &=&
\frac{1}{M} \frac{d^{2}}{d t^{2}} I_{xy}^{TT} 
\label{h-}
.
\end{eqnarray}

\begin{figure}
\resizebox{1.\textwidth}{!}{\includegraphics{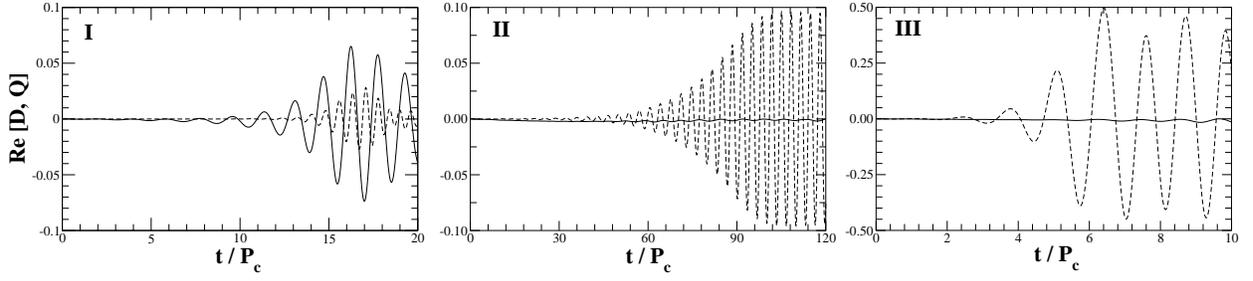}}
\caption{
Diagnostics $D$ and $Q$ as a function of $t/P_{\rm c}$ for three
differentially rotating stars (see Table~\ref{tab:initial}).  Solid
and dotted lines denote the values of $D$ and $Q$, respectively.  The
Roman numeral in each panel corresponds to the model of the
differentially rotating stars, respectively.  Hereafter $P_{\rm c}$
represents the central rotation period at $t=0$.
}
\label{fig:dig}
\end{figure}

The time evolutions of the dipole diagnostic and the quadrupole
diagnostic are plotted in Figure~\ref{fig:dig}.  We determine that the
system is stable to $m=1$ ($m=2$) mode when the dipole (quadrupole)
diagnostic remains small throughout the evolution, while the system is
unstable when the diagnostic grows exponentially at the early stage of
the evolution.  It is clearly seen in Figure~\ref{fig:dig} that the star
is more unstable to the one-armed spiral mode for model I, and more
unstable to the bar mode for models II and III.  In fact, both
diagnostics grow for model I.  The dipole diagnostic, however, grows
larger than the quadrupole diagnostic, indicating that the $m=1$ mode
is the dominant unstable mode.

\begin{figure}
\resizebox{1.\textwidth}{!}{\includegraphics{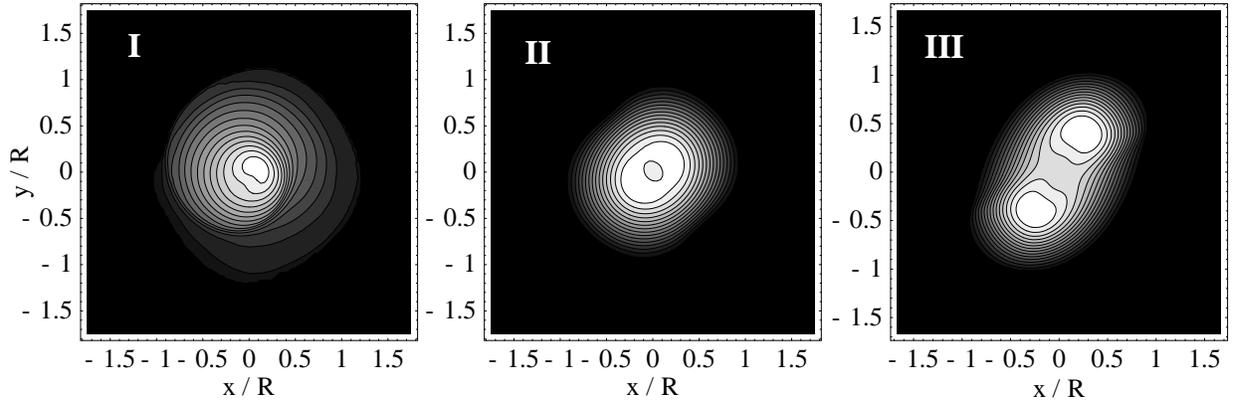}}
\caption{
Density contours in the equatorial plane for three differentially rotating
stars (see Table~\ref{tab:initial}).  Models~I, II, and III are
plotted at the parameter ($t/P_{\rm c}$, $\rho_{\rm max} / \rho_{\rm
  max}^{(0)}$) = ($16.2$, $3.63$), ($134$, $1.26$), and ($5.49$,
$1.20$), where $\rho_{\rm max}$ is the maximum density, $\rho_{\rm
  max}^{(0)}$ is the maximum density at $t=0$, and $R$ is the
equatorial radius at $t=0$.  The contour lines denote densities $\rho
/ \rho_{\rm max} = 10^{- (16-i) \times 0.287} (i=1, \cdots, 15)$ for
model~I and $\rho / \rho_{\rm max} = 6.67 (16-i) \times 10^{-2} (i=1,
\cdots, 15)$ for models~II and III, respectively.
}
\label{fig:qxy}
\end{figure}

The density contour of the differentially rotating stars are shown in 
Figure~\ref{fig:qxy} for the equatorial plane.  It is clearly seen in
Figure~\ref{fig:qxy} that one-armed spiral structure is formed at the
intermediate stage of the evolution for model I, and that bar
structure is formed for models II and III once the dynamical
instability sets in.

\begin{figure}
\resizebox{1.\textwidth}{!}{\includegraphics{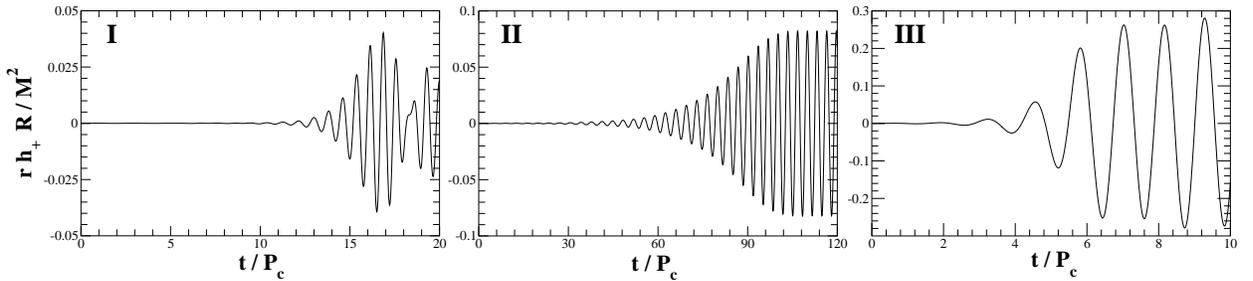}}
\caption{
Gravitational waveform for three differentially rotating stars (see
Table~\ref{tab:initial}) as seen by a distant observer located on the
rotational axis of the equilibrium star.
}
\label{fig:gw}
\end{figure}

We also show gravitational waves generated from dynamical one-armed
spiral and bar instabilities in Figure~\ref{fig:gw}.  For $m=1$ modes,
the gravitational radiation is emitted not by the primary mode itself,
but by the $m=2$ secondary harmonic which is simultaneously excited,
albeit at the lower amplitude.  Unlike the case for bar-unstable
stars, the gravitational wave signal does not persist for many
periods, but instead damp fairly rapidly.

\begin{table}
\begin{tabular}{c c c c c}
\hline
Model  &  
$\Omega_{\rm c}/\Omega_{\rm s}$  
\footnote[8]{$\Omega_{\rm s}$: Surface angular velocity} & 
$T/|W|$ & $\sigma/\Omega_{\rm c}$ 
\footnote[9]{$\sigma$: Eigenfrequency} &
$\varpi_{\rm{crt}}/\varpi_{\rm s}$ 
\footnote[10]{$\varpi_{\rm crt}$: Corotation radius; 
$\varpi_{\rm s}$: Surface radius}
\\
\hline
A-i   & 11.34 & 0.460 & $-0.245$ & ---\\
A-ii  & 11.34 & 0.460 & $0.551+0.0315 i$ & 0.281\\
A-iii & 11.34 & 0.460 & $1.15$ & ---\\
B & 13.00 & 0.170 & $0.327+0.0126 i$ & 0.507\\
\hline
\end{tabular}
\caption{Parameters for equilibrium gaseous fluid and eigenfrequency}
\label{tab:cylfrq}
\end{table}

\section{Canonical Angular Momentum to Diagnose Dynamical Instability}
\label{sec:Canonical}
We introduce the canonical angular momentum following \citet{FS78a} to 
diagnose the oscillations in rotating fluid.  In the theory of
adiabatic linear perturbations of a perfect fluid configuration with
some symmetries, it is possible to introduce canonical conserved
quantities associated with the symmetries.  Since we only use
canonical angular momentum $J_{\rm c}$ in this paper, we write down
its explicit form as
\begin{equation}
J_{\rm c} = 
  m\int_{V}(\Re[\sigma]-m\Omega)\rho|\xi|^2 dV
 -2m\int_{V} \rho \varpi\Omega\cdot \Im[\xi^\varpi\xi^{\varphi *}] dV,
\label{canonJform}
\end{equation}
where $\sigma$ is the eigenfrequency, $\xi^{i}$ is Lagrangian
displacement vector.  Note that total canonical angular momentum
becomes zero when dynamical instability sets in.

\begin{figure}
\resizebox{1.\textwidth}{!}{\includegraphics{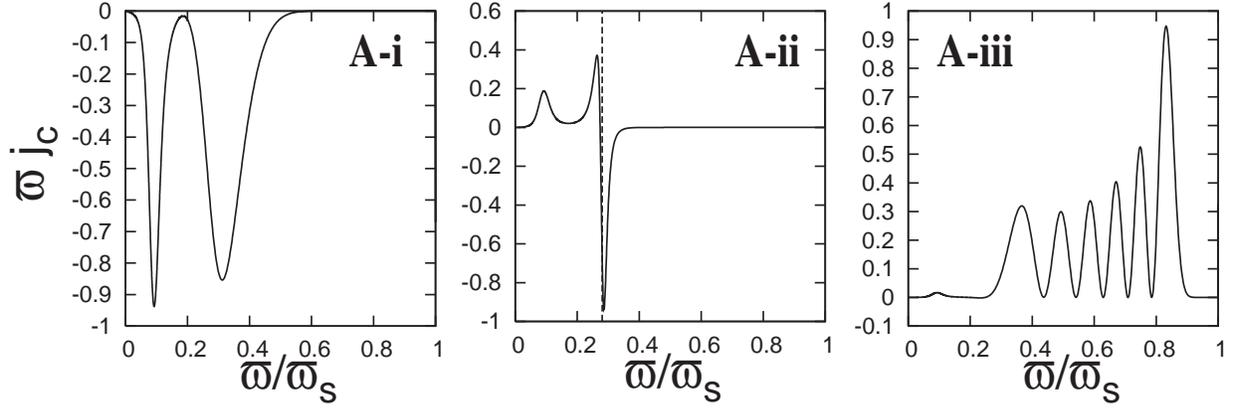}}
\caption{
Distribution of canonical angular momentum density for $m=1$ unstable
mode (see Table \ref{tab:cylfrq}).  Vertical dashed line represents
the location of corotation radius of the mode.  The Roman character in
each panel corresponds to the model of the cylindrical gaseous fluid,
respectively.  Note that we normalized the distribution of the
canonical angular momentum in an appropriate value, since the
eigenfunction can be scaled arbitrarily.
}
\label{jc-distrib-m1-stab}
\end{figure}

\begin{figure}
\resizebox{1.\textwidth}{!}{\includegraphics{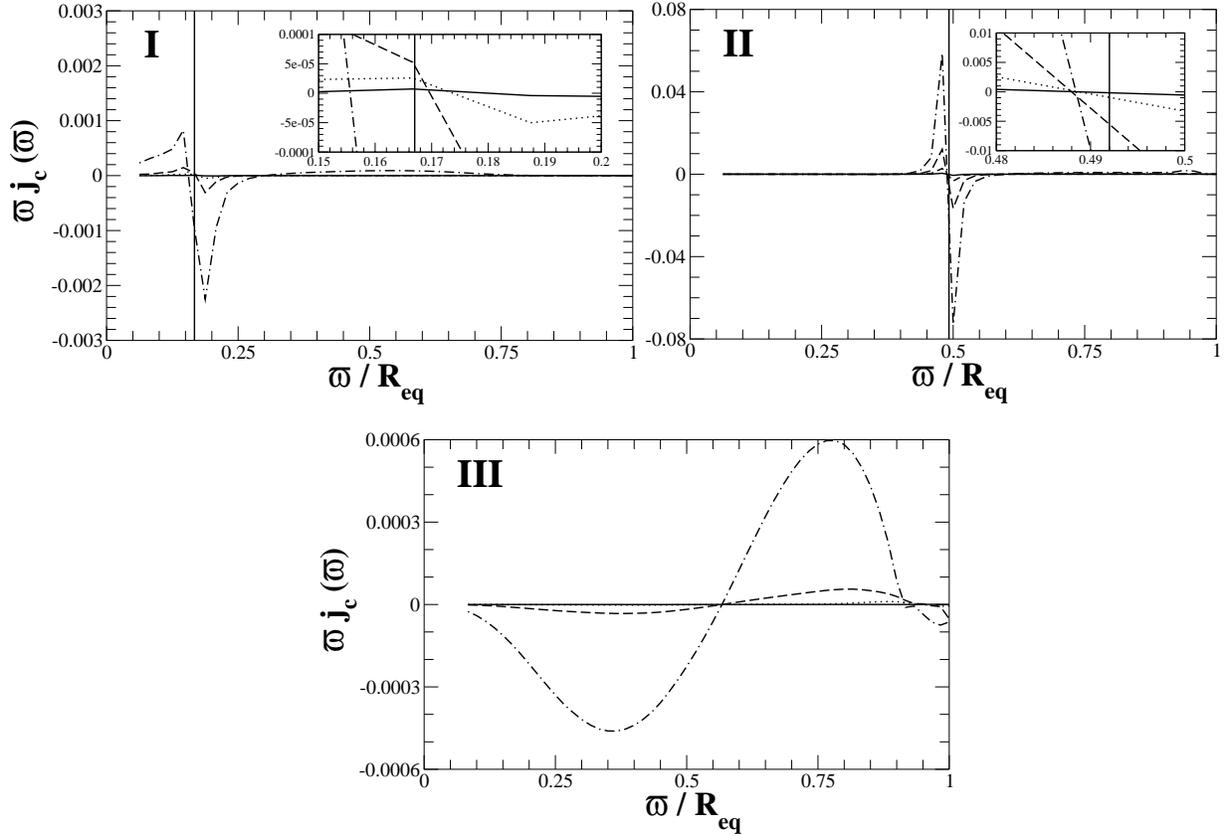}}
\caption{
Snapshots of the canonical angular momentum distribution $\varpi
j_{\rm c} (\varpi)$ in the equatorial plane for three differentially
rotating stars (see Table~\ref{tab:initial}).  Solid, dotted, dashed,
and dash-dotted line represents the time $t/P_{\rm c} =$(3.47, 6.93,
10.40, 13.86) for model~I, $t/P_{\rm c} =$(45.68, 56.43, 67.18, 77.97)
for model~II, and $t/P_{\rm c} =$(1.10, 2.19, 3.29, 4.39) for
model~III, respectively.  Note that vertical line in panels~I and II
denotes the corotation radius of the star (model~III does not have a
corotation radius).  We also enlarged the figure around the corotation
radius for panels~I and II, which is presented in the right upper part
of each panel.  Although our method of determining the corotation
radius is not precise, we clearly find that the distribution
significantly changes around the corotation radius.
}
\label{fig:jc}
\end{figure}

Next we apply the method of canonical angular momentum to the
linearized oscillations of a cylinder.  We prepare two $m=1$ stable
modes (A-i, A-iii) and one $m=1$ unstable mode (A-ii), summarized in
Table \ref{tab:cylfrq}.  We plot the integrand of canonical angular
momentum $\varpi j_{\rm c}$  
\begin{equation}
\varpi j_{\rm c}(\varpi) = m (\Re[\sigma]-m\Omega)\rho|\xi|^2
  - 2 m \rho \varpi\Omega\cdot \Im[\xi^\varpi\xi^{\varphi *}],
\end{equation}
for $m=1$ mode in Figure~\ref{fig:jc}.  We define corotation radius 
$\varpi_{\rm crt}$ of modes as $\Re[\sigma]-m\Omega(\varpi)=0$.  This
means that the pattern speed of mode coincides with the local
rotational frequency of background flow there.  Note that an integral
in the entire cylinder is zero for these cases.  The features of the
canonical angular momentum distribution for $m=1$ unstable modes are, 
\begin{enumerate}
\item 
It changes sign around corotation radius $\varpi_{\rm crt}$.
\label{ite:crt}
\item
It is positive for $\varpi<\varpi_{\rm crt}$, while 
negative for $\varpi>\varpi_{\rm crt}$. 
\label{ite:pnd}
\end{enumerate}
The feature~\ref{ite:crt} is remarkable and suggests us that the
instability is related to the corotation.  The feature has a clear
contrast for a stable mode (Figure~\ref{jc-distrib-m1-stab}).  The
canonical angular momentum is either positive or negative definite,
and it does not change its sign.  Note that the former is the case
when the pattern speed of mode is faster than the rotation of cylinder
everywhere, while the latter is the opposite.  This feature is
expected from the equation~\eqref{canonJform}, if the first term is
dominant.  In such case, the sign of $\Re[\sigma] - m \Omega(\varpi)$
determines the sign of the canonical angular momentum.  This simple
interpretation, however, does not hold for the dynamically unstable
mode.  As it is shown in the feature~\ref{ite:pnd} above, we have a
positive canonical angular momentum inside the corotation, which is
opposite to the sign of $\sigma-\Omega(\varpi)$ for
$0\le\varpi<\varpi_{\rm crt}$.

\begin{figure}
\resizebox{.4\textwidth}{!}{\includegraphics{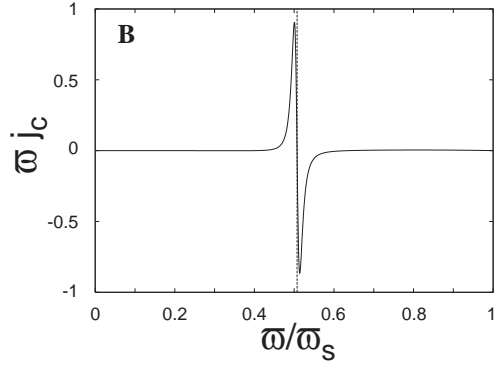}}
\caption{
Distribution of canonical angular momentum density for $m=2$ unstable
mode (see Table \ref{tab:cylfrq}).  The Roman character in the panel
corresponds to the model of the cylindrical gaseous fluid.  Vertical
dashed lines mark the locations of corotation radius of the mode.
}
\label{jc-distrib-m2-unstab}
\end{figure}

In Figure~\ref{jc-distrib-m2-unstab}, we show an example of canonical 
angular momentum distribution for $m=2$ unstable mode of
differentially rotating cylinder, which may be compared with the low
$T/|W|$ bar instability of \citet{SKE}.  We did not find $m=2$
unstable modes for the same parameters as in the case of $m=1$
instability.  The features at the corotation radius, however, are the
same as in $m=1$ instability.

It is interesting to see how the profile of the canonical angular
momentum changes when we consider the classical bar instability with
uniform rotation.  Unfortunately the bar mode of uniformly rotating
cylinder has a neutral stability point at the breakup limit.  We instead 
looked at $m=2$ instability of uniformly rotating, incompressible
Bardeen disk \citep{Bardeen75} and the classical bar instability of
Maclaurin spheroid.  These are actually more suitable for comparison
to differentially rotating spheroidal model, which we present in the
following section.  For both of the models we have analytic
expressions of oscillation modes.  It is remarkable that the canonical
angular momentum density is zero everywhere (which ensures that the
total canonical angular momentum vanishes).  This is in a clear
contrast with the $m=2$ instability in the cylinder with highly
differential rotation.

\begin{table}
\begin{tabular}{ccc}
\hline
Model &
$\sigma$ $[\Omega_{\rm c}]$ &
$\varpi_{\rm crt}$  $[R_{\rm eq}]$
\\
\hline
I & $0.590 + 0.0896 i$ & $0.167$
\\
II & $0.284 + 0.0121 i$ & $0.492$ 
\\
III & $0.757 + 0.200 i$ & ---
\\
\hline
\end{tabular}
\caption{Eigenfrequency and the corotation radius of three differentially
  rotating stars}
\label{tab:freq}
\end{table}

Finally, we adopt the method of canonical angular momentum to the
nonlinear hydrodynamics.  We identify the complex eigenfrequency and
the corotation radius from dipole or quadrupole diagnostics which is
summarized in Table \ref{tab:freq}.  Note that we read off the
eigenfrequency from those plots at the early stage of the evolution.
The Eulerian perturbed velocity is defined by subtracting the
velocity at equilibrium from the velocity.  The Lagrangian
displacement vector is extracted by using a linear formula for a
dominant mode in each case.

We show the snapshots of canonical angular momentum density in
Figure~\ref{fig:jc}.  Since we determine the corotation radius using
the extracted eigenfrequency and the angular velocity profile at
equilibrium, the radius does not change throughout the evolution.  For
low $T/|W|$ dynamical instability, the distribution of the canonical
angular momentum drastically changes its sign around the corotation
radius, and the maximum amount of canonical angular momentum density
increases at the early stage of evolution.  This means that the
angular momentum flows inside the corotation radius in the
evolution.  On the other hand, for high $T/|W|$ dynamical instability
(Panel~III of Figure~\ref{fig:jc}), which may be regarded as a
classical bar instability modified by differential rotation, the
distribution of the canonical angular momentum is smooth and with no
particular feature.

Note that the amplitude of $\varpi j_c$ is orders of magnitude smaller
than those in the corotating cases in top and middle panels of
Figure~\ref{fig:jc}.  Contrary to the linear perturbation analysis,
the amplitude here is not scale free and the relative amplitude has a
physical meaning.  Thus the smallness of it for model~III suggest that
it should be exactly zero everywhere in the limit of linearized
oscillation.  The deviation from zero may come from the imperfect
assumption of linearized oscillation, that is made here to extract
oscillation frequency and Lagrangian displacement vector.

From these different behaviours of the distribution of the canonical
angular momentum, we find that the mechanisms working in the low
$T/|W|$ instabilities and the classical bar instability may be quite
different, i.e., in the former the corotation resonance of modes are
essential, while the instability is global in the latter case.

\section{Summary and Discussion}
\label{sec:Discussion}

We have studied the nature of three different types of dynamical
instability in differentially rotating stars both in linear eigenmode
analysis and in hydrodynamic simulation using canonical angular
momentum distribution.

We have found that the low $T/|W|$ dynamical instability occurs around
the corotation radius of the star by investigating the distribution of
the canonical angular momentum.  We have also found by investigating
the canonical angular momentum that the instability grows through the
inflow of the angular momentum inside the corotation radius.  The
feature also holds for the dynamical bar instability in low $T/|W|$,
which is in clear contrast to that of classical dynamical bar
instability in high $T/|W|$.  Therefore the existence of corotation
point inside the star plays a significant role of exciting one-armed
spiral mode and bar mode dynamically in low $T/|W|$.  However, we made
our statement from the behaviour of the canonical angular momentum,
the statement holds only in a sense of necessary condition.  In order
to understand the physical mechanism of the low $T/|W|$ dynamical
instability, we need another tool and it will be the next step of this
study.

The feature of gravitational waves generated from these instabilities
are also compared.  Quasi-periodic gravitational waves emitted by
stars with $m=1$ instabilities have smaller amplitudes than those
emitted by stars unstable to the $m=2$ bar mode.  For $m=1$ modes, the
gravitational radiation is emitted not directly by the primary mode
itself, but by the $m=2$ secondary harmonic which is simultaneously
excited.  Possibly this $m=2$ oscillation is generated through a
quadratic nonlinear selfcoupling of $m=1$ eigenmode.  Remarkably the
precedent studies \citep{CNLB01,SBS03} found that the pattern speed of
$m=2$ mode is almost the same as that of $m=1$ mode, which suggest the
former is the harmonic of the latter.  Unlike the case for
bar-unstable stars, the gravitational wave signal does not persist for
many periods, but instead is damped fairly rapidly.  We have not
understood this remarkable difference between $m=1$ and $m=2$ unstable
cases.  One of the possibility may be that the unstable $m=1$
eigenmode tends to couple to higher and higher $m$ modes (which are
not necessarily unstable and could be in the continuous spectrum) and
pump its energy to them in a cascade way.  However, we have not found
the feature that prevents $m=2$ mode from this cascade dissipation.

Another possibility is that the spiral pattern formed in $m=1$
instability redistributes the angular momentum of the original
unstable flow, so that the flow is quickly stabilized.  Inside the
corotation radius, the background flow is faster than the pattern,
while it is slower outside.  A similar mechanism to Landau damping in
plasma wave which transfer the momentum of wave to background flow may
work at the spiral pattern.  The pattern may decelerate the background
flow inside the corotation and accelerate it outside the corotation,
which may change the unstable flow profile to stable one.  As we do
not see a spiral pattern forming in the low $T/|W|$ bar instability,
it may eludes this damping process.


%
\end{document}